\documentclass[aps,prd,preprint,superscriptaddress,amsmath,amssymb,showpacs,nofootinbib]{revtex4-1}
\usepackage{dcolumn}
\usepackage{graphicx}
\usepackage{mathrsfs}
\usepackage{float}
\usepackage{physics}
\usepackage{xcolor}
\usepackage[colorlinks=true,allcolors=blue]{hyperref}

\newcommand{\e}{\mathrm{e}}
\newcommand{\CP}{\mathbb{CP}}

\begin{document}
\title{Projective Origin of the Spin Hydrodynamic Attractor and Its Resurgent Repellor}

\author{Qi Zhou}
\email{qizhou@m.scnu.edu.cn}
\address{State Key Laboratory of Nuclear Physics and Technology, Institute of Quantum Matter, South China Normal University, Guangzhou 510006, China}
\address{Guangdong Basic Research Center of Excellence for Structure and Fundamental Interactions of Matter, Guangdong Provincial Key Laboratory of Nuclear Science, Guangzhou 510006, China}

\author{Enke Wang}
\email{wangek@scnu.edu.cn}
\address{State Key Laboratory of Nuclear Physics and Technology, Institute of Quantum Matter, South China Normal University, Guangzhou 510006, China}
\address{Guangdong Basic Research Center of Excellence for Structure and Fundamental Interactions of Matter, Guangdong Provincial Key Laboratory of Nuclear Science, Guangzhou 510006, China}

\begin{abstract}
We investigate the projective and resurgent structure of a spin hydrodynamic attractor in Bjorken expansion. 
We show that the nonlinear spin solution family is determined by the projective classes of the two dimensional linear solution space, with the attractor and repellor corresponding to two distinguished projective directions and the linear modes ratio generating the full one-parameter transseries tower.
Using the projective transseries structure, we show analytically that the data of repellor are encoded in the Borel-Stokes structure of the attractor expansion.
These results provide an explicit analytic realization of resurgence relations that are often extracted through high order expansions and numerical Borel analysis, and yield a unified description of the attractor, the repellor, and their Borel-Stokes connection in minimal causal spin hydrodynamics.
\end{abstract}

\maketitle

\section{Introduction}
\label{sec:intro}

In non-central relativistic heavy-ion collisions, the large orbital angular momentum deposited in the interaction region generates substantial vortical motion in the quark--gluon plasma (QGP), which can polarize microscopic spin degrees of freedom through spin--orbit coupling~\cite{Liang:2004ph,Liang:2004xn,Becattini:2013fla}. 
Experimental evidence for spin-polarization phenomena includes measurements of global $\Lambda$ polarization~\cite{STAR:2017ckg}, multistrange-hyperon polarization~\cite{STAR:2020xbm}, and vector-meson spin alignment~\cite{STAR:2022fan}. 
Nevertheless, several issues remain unresolved, including particle--antiparticle differences in hyperon polarization~\cite{STAR:2023nvo,Peng:2022cya}, the origin of local hyperon polarization~\cite{Karpenko:2016jyx}, and the flavor dependence of vector-meson spin alignment~\cite{ALICE:2022dyy}; see Refs.~\cite{Gao:2020vbh,Becattini:2020ngo,Becattini:2022zvf,Hidaka:2022dmn,Becattini:2024uha,Shen:2020mgh} for reviews. Relativistic spin hydrodynamics extends conventional hydrodynamics by incorporating spin degrees of freedom while consistently implementing energy--momentum and total-angular-momentum conservation. 
Its formulation has been developed from kinetic theory~\cite{Florkowski:2017ruc,Florkowski:2018fap,Bhadury:2020puc,Weickgenannt:2022qvh}, entropy-current analysis~\cite{Hattori:2019lfp,Fukushima:2020ucl,Hongo:2021ona}, effective hydrodynamic theories and holography~\cite{Gallegos:2020otk,Gallegos:2021bzp}, and the non-equilibrium statistical-operator formalism~\cite{Hu:2021lnx,Tiwari:2024trl,She:2024rnx}. These developments provide a natural framework for investigating whether spin evolution exhibits universal far-from-equilibrium behavior analogous to hydrodynamic attractors.

The hydrodynamic attractor shows that hydrodynamization need not require convergence of the gradient expansion or complete local equilibration. 
Instead, transient non-hydrodynamic modes decay, causing solutions with different initial conditions to approach a universal trajectory~\cite{Heller:2015dha}. 
Such attractors have been identified in kinetic theory, viscous and anisotropic hydrodynamics, and strongly coupled holographic systems~\cite{Heller:2016rtz,Denicol:2016bjh,Romatschke:2017vte,Strickland:2018ayk}. 
The hydrodynamic attractor admits a fixed-point interpretation as a special trajectory connecting the early-time collisionless fixed point to the late time hydrodynamic fixed point~\cite{Blaizot:2017ucy,Blaizot:2019scw,Blaizot:2021cdv}.
Recent developments have related attractor behavior to dynamical dimensionality reduction in phase space and, in kinetic theory, to adiabatic hydrodynamization governed by slowly evolving pre-hydrodynamic modes~\cite{Heller:2020anv,Spalinski:2025ngd,Brewer:2019oha,Rajagopal:2024lou}.
Resurgence further shows that the divergent hydrodynamic gradient expansion encodes information about transient non-hydrodynamic sectors, with transseries relating the hydrodynamic and non-hydrodynamic contributions~\cite{Heller:2015dha,Basar:2015ava,Aniceto:2015mto,Heller:2016rtz,Aniceto:2018uik}.
Extending these ideas to spin hydrodynamics therefore motivates the search for spin attractors and for a resurgent structure that systematically characterizes both hydrodynamic and non-hydrodynamic spin dynamics.

Recent studies of Bjorken spin hydrodynamics have identified attractor-like behavior in nonlinear variables constructed from spin density modes~\cite{Wang:2024afv,Zhou:2026jbh}. 
An important feature of these formulations is that the underlying spin density mode satisfies a second order linear differential equation, whereas the corresponding attractor variable is defined through a nonlinear logarithmic-derivative map.
Closely related linear-to-nonlinear reductions and analytic solution methods have appeared in earlier studies of Bjorken flow attractors~\cite{Denicol:2017lxn,Denicol:2019lio,Jaiswal:2019cju,Blaizot:2020gql,Blaizot:2021cdv}.
Among these, Blaizot and Yan showed explicitly how a nonlinear attractor equation can emerge from an underlying linear two mode system~\cite{Blaizot:2021cdv}, while related analytic structures are present in the exact solution approaches of Denicol and Noronha~\cite{Denicol:2017lxn,Denicol:2019lio} and Jaiswal et al.~\cite{Jaiswal:2019cju}, as well as in a kinetic theory moment framework~\cite{Blaizot:2020gql,Aniceto:2024pyc}.
Following the strategy developed in Refs.~\cite{Heller:2015dha,Basar:2015ava}, we adopt a late time transseries ansatz for the fundamental solutions and derive their exact Borel transforms.
The exact solvability of the spin density equation further allows the attractor to be selected by early-time regularity, while its late time structure can be analyzed within a canonical transseries framework.
Motivated by these developments, we formulate the spin sector dynamics in a projective framework. 
The two dimensional linear solution space of the spin density equation induces a one-parameter family of nonlinear solutions, with the attractor and repellor corresponding to distinguished projective rays. 
This perspective clarifies which features of the spin attractor are inherited from the underlying linear dynamics and provides a direct geometric relation between the attractor and repellor.

The attractor repellor structure provides a broader dynamical-systems perspective on the standard hydrodynamic attractor picture~\cite{Behtash:2019qtk,Jaiswal:2019cju,Jaiswal:2021uvv}. 
In Bjorken flow dynamics, the repellor may correspond to unstable fixed points or fixed lines associated with early time evolution~\cite{Jaiswal:2021uvv}.  
Such a phase-space organization has been made explicit in analytic and exact solution studies of higher order viscous hydrodynamics, where generic attracting solutions and special non-attracting branches are distinguished by transient sectors and integration constants~\cite{Denicol:2017lxn,Jaiswal:2019cju}.
Related attractor repellor structures have been studied beyond conformal viscous hydrodynamics, including in nonconformal kinetic theory~\cite{Jaiswal:2021uvv,Chen:2021wwh}.
However, to our knowledge, whether the attractor and repellor branches are directly related at the level of asymptotic and Borel--Stokes data has not been clarified.
In the spin density sector, this question can be analyzed particularly clearly since both branches originate from the same second order linear equation, whose two independent solutions have direct physical meaning as spin density modes~\cite{Wang:2024afv,Zhou:2026jbh}.
These observations support treating the attractor and repellor as complementary branches of a common linear and geometric structure, and motivate investigating whether they are also related through resurgence.
Building on these framework, we further show that the resurgence structure of the attractor branch encodes data of the repellor.

The paper is organized as follows.  
In Sec.~\ref{sec:projective_spin} we establish the projective structure of the spin density solution space, identify the attractor and repellor, and construct the corresponding late time transseries.  
In Sec.~\ref{sec:spin_resurgence} we study the corresponding Borel-plane structure and show that the Stokes discontinuity of the attractor asymptotic expansion encodes the contribution associated with the repellor.  
We conclude in Sec.~\ref{sec:conclusion}.

\section{Projective origin of the spin attractor}
\label{sec:projective_spin}

Motivated by recent studies of spin hydrodynamic attractors in Bjorken expansion~\cite{Wang:2024afv,Zhou:2026jbh}, and by earlier analytic studies revealing closely related linear or linearizable structures underlying nonlinear attractor dynamics~\cite{Denicol:2017lxn,Denicol:2019lio,Jaiswal:2019cju,Blaizot:2020gql,Blaizot:2021cdv}, we investigate the projective origin of the spin attractor.
In particular, the logarithmic-derivative construction of Ref.~\cite{Blaizot:2021cdv} can be viewed as implicitly involving the projectivization of the full two-dimensional linear solution space, although the underlying projective geometry was not formulated explicitly. 
Here we formulate this structure directly and identify the attractor, repellor, and the complete one-parameter nonlinear solution family with distinguished points and trajectories in the associated projective space.
We establish the projective structure of the exact linear solution space and identify the distinguished attractor and repellor solutions. We then use this structure to construct the closed late time transseries of the nonlinear spin solution.

\subsection{Projective Structure, Attractor and Repellor}
Using $w$ to denote the dimensionless time variable, we consider a Bjorken $(0+1)$D spin density mode $\mathcal{S}(w)$ governed by a second order linear equation 
\begin{equation}
    \mathcal{L}\,\mathcal{S}(w)=0 .
    \label{eq:linearSpinEquation}
\end{equation}
Here, the linear differential operator is explicitly given by
\begin{equation}
    \mathcal{L}
    =
    \frac{d^2}{dw^2}
    +
    \left(\frac{3}{2}+\frac{1}{w}\right)\frac{d}{dw}
    +
    9\lambda+\frac{9}{4w}-\frac{9}{4w^2},
\end{equation}
which arises in minimal causal spin hydrodynamics in conformal limit~\cite{Wang:2024afv,Zhou:2026jbh}.

Before introducing the late time expansion, it is useful to exploit the fact that Eq.~\eqref{eq:linearSpinEquation} is exactly solvable.
Similar exact solution and linearization methods have been employed in studies of hydrodynamic attractors in Bjorken flow~\cite{Blaizot:2021cdv,Denicol:2017lxn,Denicol:2019lio,Jaiswal:2019cju}. We define
\begin{equation}
    \Delta=\sqrt{1-16\lambda},
    \qquad
    \Xi={3\over2}\Delta,
    \qquad
    \kappa={1\over\Delta},
    \qquad
    z=\Xi w,
    \label{eq:WhittakerVariables}
\end{equation}
and introduce
\begin{equation}
    \mathcal{S}(w)=\e^{-3w/4}w^{-1/2}Y(z).
    \label{eq:WhittakerTransform}
\end{equation}
The linear equation then becomes
\begin{equation}
    {d^2Y\over dz^2}
    +
    \left[
        -{1\over4}
        +{\kappa\over z}
        +{1/4-\mu^2\over z^2}
    \right]Y=0,
    \qquad
    \mu={3\over2},
    \label{eq:WhittakerEquation}
\end{equation}
which is the Whittaker equation. 
We focus on $0<\lambda<{1\over16}$ for simplicity, so that $\Delta$ is real and positive~\cite{Zhou:2026jbh}. The free parameter $\lambda$ originates from the source term in the IS-type relaxation equation of the rotational tensor, while the limit $\lambda\to0$ corresponds to the homogeneous limit of spin hydrodynamics.

Analytic exact solution studies have used fixed point and asymptotic conditions to identify distinguished attractor solutions in Bjorken flow dynamics~\cite{Blaizot:2021cdv,Denicol:2017lxn,Denicol:2019lio,Jaiswal:2019cju}. 
In the present formulation, we extend this idea to the underlying second order linear equation, for which regularity at the singular point $w=0$ selects a distinguished projective ray. 
Performing a dominant balance analysis near $w=0$, Eq.~\eqref{eq:linearSpinEquation} reduces to
\begin{equation}
    \mathcal{S}''+{1\over w}\mathcal{S}'-{9\over4w^2}\mathcal{S}\simeq0,
\end{equation}
whose two independent local behaviors are
\begin{equation}
    \mathcal{S}(w)\sim w^{3/2},
    \qquad
    \mathcal{S}(w)\sim w^{-3/2}.
    \label{eq:earlyFrobenius}
\end{equation}
We therefore select the regular solution and identify it with the attractor linear mode. 
For generic nonresonant values $\kappa\notin\{2,3,4,\ldots\}$, it is convenient to normalize the two exact solutions as
\begin{align}
    \mathcal{S}_+(w)
    &=
    {\Gamma(2-\kappa)\over6}\,\Xi^\kappa
    \e^{-3w/4}w^{-1/2}
    M_{\kappa,3/2}(\Xi w),
    \label{eq:ExactRegularSplus}
    \\
    \mathcal{S}_-(w)
    &=
    \Xi^{-\kappa}
    \e^{-3w/4}w^{-1/2}
    W_{\kappa,3/2}(\Xi w),
    \label{eq:ExactSubdominantSminus}
\end{align}
where $M_{\kappa,\mu}(z)$ and $W_{\kappa,\mu}(z)$ denote the Whittaker $M$ and $W$ functions, respectively.
The normalization of $\mathcal{S}_+$ is chosen such that its dominant late time asymptotic coefficient is unity. 
In the early time limit $w\to0$, these solutions behave as
\begin{equation}
    \mathcal{S}_+(w)\propto w^{3/2}[1+O(w)],
    \qquad
    \mathcal{S}_-(w)\propto w^{-3/2}[1+O(w)],
    \label{eq:earlyExactModes}
\end{equation}
so that the corresponding logarithmic derivatives satisfy
\begin{equation}
    f_+(w)\to1,
    \qquad
    f_-(w)\to-1,
    \qquad
    w\to0.
    \label{eq:earlyFixedValues}
\end{equation}
For a detailed analysis and the corresponding attractor flow lines, see Ref.~\cite{Zhou:2026jbh}.
At the isolated polynomial values $\kappa=2,3,\ldots$, the Whittaker basis above becomes degenerate and the dominant coefficient of the regular solution vanishes. These values require a separate limiting basis and will not be considered in the following generic discussion.

The nonlinear spin attractor variable is defined through the logarithmic derivative~\cite{Wang:2024afv,Zhou:2026jbh}
\begin{equation}
    f(w)={2w\over3}{\mathcal{S}'(w)\over \mathcal{S}(w)} .
    \label{eq:fDef}
\end{equation}
Since Eq.~\eqref{eq:linearSpinEquation} is linear, a generic solution can be written in the exact regular basis as
\begin{align}
    \mathcal{S}(w)
    &=A \mathcal{S}_+(w)+B \mathcal{S}_-(w)
    =A \mathcal{S}_+(w)\left[1+\sigma \mathcal{R}(w)\right],
    \label{eq:linearSuperposition}
    \\
    \sigma&={B\over A},
    \qquad
    \mathcal{R}(w)={\mathcal{S}_-(w)\over \mathcal{S}_+(w)}.
    \nonumber
\end{align}
The regularity condition in Eq.~\eqref{eq:earlyFrobenius} eliminates the singular contribution and therefore fixes
\begin{equation}
    B=0,
    \qquad
    \sigma=0.
    \label{eq:regularitySigmaZero}
\end{equation}
Thus, once the basis in Eqs.~\eqref{eq:ExactRegularSplus} and \eqref{eq:ExactSubdominantSminus} is chosen, $\mathcal{S}_+$ is not merely a dominant asymptotic branch. It is the unique regular projective ray, up to an irrelevant overall normalization.

The overall normalization $A$ drops out of Eq.~\eqref{eq:fDef}. The nonlinear solution therefore depends only on the projective coordinate $\sigma=B/A$,
\begin{equation}
    f_\sigma(w)
    ={2w\over3}{\mathrm{d}\over\mathrm{d} w}
    \log\!\left(\mathcal{S}_+(w)+\sigma \mathcal{S}_-(w)\right).
    \label{eq:projectiveFamily}
\end{equation}
Equivalently,
\begin{equation}
    f_\sigma(w)
    =f_+(w)+{2w\over3}{\sigma \mathcal{R}'(w)\over1+\sigma \mathcal{R}(w)},
    \qquad
    f_\pm(w)={2w\over3}{\mathcal{S}_\pm'(w)\over \mathcal{S}_\pm(w)} .
    \label{eq:fSigma}
\end{equation}
The map
\begin{equation}
    (A,B)\in\mathbb{C}^2\setminus\{0\}
    \quad\longrightarrow\quad
    [A:B]\in\CP^1
    \quad\longrightarrow\quad
    f_\sigma(w)
    \label{eq:projectiveMap}
\end{equation}
is the projective origin of the nonlinear spin solution family. The physical real solutions form the real slice $\mathbb{RP}^1\subset\mathbb{CP}^1$. In the regular basis, the attractor is the distinguished projective ray $[1:0]$, or equivalently $\sigma=0$, whereas the pure singular/subdominant solution is the ray $[0:1]$.
The nonlinear solution $f_-$ associated with the pure $\mathcal{S}_-$ branch is identified as the repellor, since any nonzero admixture of $\mathcal{S}_+$ grows relative to $\mathcal{S}_-$ at large $w$ and drives the solution away from $f_-$.

\subsection{Late time Transseries Structure}

As a direct application of the projective structure established above, we now construct the closed late time transseries representation of the nonlinear spin solution.
The two formal late time sectors take the form
\begin{equation}
    S_\pm(w)
    \sim
    \e^{r_\pm w}w^{p_\pm}\Phi_\pm(w),
    \qquad
    \Phi_\pm(w)\sim\sum_{n=0}^{\infty}{c_n^{(\pm)}\over w^n},
    \qquad
    c_0^{(\pm)}=1,
    \label{eq:Ssectors}
\end{equation}
where
\begin{equation}
    r_\pm=-{3\over4}\pm{3\over4}\Delta,
    \qquad
    p_\pm=-{1\over2}\mp{1\over\Delta}.
    \label{eq:rpm}
\end{equation}
The exact Whittaker solution shows explicitly that the formal series in Eq.~\eqref{eq:Ssectors} are asymptotic expansions of actual solutions. For $\mathcal{S}_+$, Eq.~\eqref{eq:Ssectors} denotes its dominant Poincar\'e asymptotic sector; an exponentially small subdominant component fixed by regularity is invisible at every algebraic order and will be made explicit in the Stokes analysis below.

Since $S_+$ is less damped at late time while $S_-$ is subdominant, their exponential separation is
\begin{equation}
    \Xi\equiv r_+-r_-={3\over2}\Delta>0,
    \label{eq:XiDef}
\end{equation}
and
\begin{equation}
    R(w)={S_-(w)\over S_+(w)}
    \sim
    \e^{-\Xi w}w^{p_- - p_+}
    \left[1+O(w^{-1})\right],
    \qquad
    w\to\infty.
    \label{eq:ratioS}
\end{equation}
It follows from Eq.~\eqref{eq:fSigma} that
\begin{equation}
    f_\sigma(w)-f_+(w)
    \sim
    {2w\over3}\sigma
    \left(-\Xi+{\beta\over w}\right)
    \e^{-\Xi w}w^\beta+\cdots,
    \qquad
    \beta=p_- - p_+={2\over\Delta}.
    \label{eq:attraction}
\end{equation}
Therefore all solutions with finite $\sigma$ approach the unique regular solution $f_+$ at late time. The solution $f_+$ is consequently selected in two complementary ways: it is the regular projective ray at $w\to0$, and it is the attracting solution for generic finite projective coordinates at $w\to\infty$. The pure repellor requires the fine-tuned ray $A=0$, for which $\mathcal{S}=\mathcal{S}_-$ and $f=f_-$. Any nonzero admixture of $\mathcal{S}_+$ grows relative to $\mathcal{S}_-$ as $\e^{\Xi w}w^{-\beta}$.

The regular basis is most convenient for identifying the exact attractor, whereas the canonical late time basis is more convenient for organizing the transseries and its Stokes phenomenon. 
To separate these two roles, we introduce the two lateral dominant Whittaker solutions
\begin{equation}
    \mathcal{S}_{\rm dom}^{(\pm)}(w)
    \equiv
    \Xi^\kappa
    \e^{\pm i\pi\kappa}
    \e^{-3w/4}w^{-1/2}
    W_{-\kappa,3/2}\!\left(\e^{\pm i\pi}\Xi w\right).
    \label{eq:CanonicalDominantWhittaker}
\end{equation}
Each $\mathcal{S}_{\rm dom}^{(\pm)}$ is an exact lateral solution whose dominant late time asymptotic expansion coincides with the formal sector $S_+$ introduced above,
\begin{equation}
    \mathcal{S}_{\rm dom}^{(\pm)}(w)
    \sim
    \e^{r_+ w}w^{p_+}\Phi_+(w)
    =
    S_+(w),
    \qquad
    w\to\infty.
    \label{eq:DominantFormalSector}
\end{equation}
Thus, $\mathcal{S}_{\rm dom}^{(\pm)}$ provides the two lateral exact realizations of the same dominant formal late time sector $S_+$.
For either lateral choice, the exact regular solution can be written as
\begin{equation}
    \mathcal{S}_+(w)=\mathcal{S}_{\rm dom}^{(\pm)}(w)+\rho_{\rm att}^{(\pm)}\mathcal{S}_-(w),
    \label{eq:RegularAsShiftedDominant}
\end{equation}
where $\rho_{\rm att}^{(\pm)}$ is a fixed number determined by regularity. Its explicit value will be derived in Sec.~\ref{sec:whittaker_completion}.

A generic solution written in the regular projective coordinate $\sigma$ can therefore also be written as
\begin{equation}
    \mathcal{S}(w)
    =A\left[\mathcal{S}_{\rm dom}^{(\pm)}(w)+\rho_\pm \mathcal{S}_-(w)\right],
    \qquad
    \rho_\pm=\rho_{\rm att}^{(\pm)}+\sigma.
    \label{eq:rhoSigmaShift}
\end{equation}
Thus $\sigma=0$ selects the exact regular attractor, while $\rho_\pm$ is the canonical late time transseries coordinate. The two coordinates differ only by a fixed projective shift.

Using the same construction as in the previous subsection, we define
\begin{equation}
    \mathcal{R}_{\rm dom}^{(\pm)}(w)={\mathcal{S}_-(w)\over \mathcal{S}_{\rm dom}^{(\pm)}(w)},
\end{equation}
the corresponding nonlinear solution is
\begin{equation}
    f_{\rho_\pm}(w)
    =f_{\mathcal{S}_{\rm dom}^{(\pm)}}(w)
    +{2w\over3}{\rho_\pm \mathcal{R}_{\rm dom}^{(\pm)\prime}(w)\over1+\rho_\pm \mathcal{R}_{\rm dom}^{(\pm)}(w)},
    \qquad
    f_{\mathcal{S}_{\rm dom}^{(\pm)}}(w)={2w\over3}{\mathcal{S}_{\rm dom}^{(\pm)\prime}(w)\over \mathcal{S}_{\rm dom}^{(\pm)}(w)}.
    \label{eq:canonicalProjectiveFamily}
\end{equation}
At late time,
\begin{equation}
    \mathcal{R}_{\rm dom}^{(\pm)}(w)
    \sim
    \e^{-\Xi w}w^\beta\Psi(w),
    \qquad
    \Psi(w)
    \sim
    1+\sum_{k=1}^{\infty}{\psi_k\over w^k},
    \label{eq:QWithPsi}
\end{equation}
where the formal series $\Psi$ is determined by the ratio of the fluctuation series of the two canonical asymptotic sectors.

For fixed finite $\rho_\pm$, expanding Eq.~\eqref{eq:canonicalProjectiveFamily} gives
\begin{equation}
    f_{\rho_\pm}(w)
    =
    f_{\mathcal{S}_{\rm dom}^{(\pm)}}(w)
    +
    {2w\over3}
    \sum_{n=1}^{\infty}
    (-1)^{n-1}\rho_\pm^n
    \left[\mathcal{R}_{\rm dom}^{(\pm)}(w)\right]^{n-1}\mathcal{R}_{\rm dom}^{(\pm)\prime}(w).
    \label{eq:fInfiniteExpansion}
\end{equation}
Using Eq.~\eqref{eq:QWithPsi}, the corresponding formal transseries takes the form
\begin{equation}
    f_{\rho_\pm}(w)
    \sim
    f_{\rm dom}^{(0)}(w)
    +
    \sum_{n=1}^{\infty}
    \rho_\pm^n
    \e^{-n\Xi w}
    w^{n\beta+1}
    \Phi_n^{(f)}(w),
    \label{eq:fFullTransseries}
\end{equation}
where $f_{\rm dom}^{(0)}$ denotes the common formal perturbative expansion of the two lateral functions $f_{\mathcal{S}_{\rm dom}^{(\pm)}}$, and
\begin{equation}
    \Phi_n^{(f)}(w)
    =
    (-1)^{n-1}{2\over3}
    \Psi^n(w)
    \left[
        -\Xi+{\beta\over w}
        +{\Psi'(w)\over\Psi(w)}
    \right]
    \sim
    \sum_{k=0}^{\infty}{f_{n,k}\over w^k}.
    \label{eq:fSectorFluctuation}
\end{equation}
The leading coefficient is
\begin{equation}
    f_{n,0}={2\Xi\over3}(-1)^n.
\end{equation}
Although the underlying linear equation contains only two independent modes, the logarithmic-derivative map generates an infinite tower of nonlinear transseries sectors with actions
\begin{equation}
    \mathcal{A}_n=n\Xi,
    \qquad
    n=1,2,\ldots.
    \label{eq:spinActions}
\end{equation}
These sectors are organized by a single canonical projective coordinate $\rho_\pm$. The regularity condition does not remove this transseries structure. Instead, it fixes the value of the transseries parameter to $\rho_\pm=\rho_{\rm att}^{(\pm)}$ for the attractor, or equivalently $\sigma=0$ in the regular basis.
The distinction between $\sigma$ and $\rho_\pm$ is useful in the resurgent analysis below. The coordinate $\sigma$ is adapted to the exact regular attractor and vanishes on it. The coordinate $\rho_\pm$ is adapted to the canonical late time Stokes basis and makes the transseries tower manifest. Their fixed shift encodes the beyond-all-orders information that is invisible in the algebraic late time expansion.

Our analysis within the minimal causal spin-hydrodynamics framework makes the underlying projective structure explicit by identifying the nonlinear solution family with the projectivization of the two-dimensional linear solution space, with the attractor and repellor corresponding to attracting and repelling projective directions.
The projective formulation makes this origin more transparent by showing directly how the logarithmic-derivative map of two linear modes generates the full nonlinear transseries tower.
Building on Ref.~\cite{Blaizot:2021cdv}, we identify the attractor and repellor as distinguished solution branches associated with the two projective directions of the underlying linear solution space.

One key difference lies in how the transseries is organized. In Ref.~\cite{Blaizot:2021cdv}, the zeroth transseries sector is the hydrodynamic gradient expansion, while the attractor is reconstructed from the full transseries after nonperturbative completion and resummation. In our projective formulation, the zeroth sector is instead associated directly with the attractor branch, while the higher sectors describe admixtures of the repelling linear mode, as in the special case discussed in Ref.~\cite{Denicol:2017lxn}.

\section{Resurgent Structure and Repellor Data In the Borel Plane}
\label{sec:spin_resurgence}

Having established the exact projective structure and its canonical late time transseries representation, we now turn to the resurgent relation between the attractor and repellor solutions. The central question is whether the information associated with the repellor is already encoded in the late time expansion of the attractor. We show that this relation is realized through the Borel--Stokes structure of the attractor expansion: its leading Borel singularity is associated with the exponential separation between the two underlying linear modes, while the corresponding Stokes discontinuity generates the contribution associated with the repellor. The projective structure derived above then allows this nonperturbative contribution to be related directly to the repellor solution. 
We further verify the same Stokes connection independently from the exact Borel transform and the exact Whittaker solution of the underlying linear equation.

\subsection{Borel singularities and large order behavior}

The exact attractor $f_+(w)$ and the canonical dominant functions $f_{\mathcal{S}_{\rm dom}^{(\pm)}}(w)$ share the same algebraic late time expansion. We denote this formal perturbative sector by
\begin{equation}
    f_{\rm dom}^{(0)}(w)
    \sim
    {\Delta-1\over2}w
    -{2+\Delta\over3\Delta}
    +\sum_{k=1}^{\infty}{a_k\over w^k}.
    \label{eq:fPerturbativeSector}
\end{equation}
The difference between the exact regular attractor and a particular lateral realization of this formal series is exponentially small and is fixed by the regularity condition. It therefore does not alter the algebraic coefficients $a_k$ or their Borel singularity structure.

Since the polynomial terms do not affect the singularity structure away from the origin of the Borel plane, we define
\begin{equation}
    \widehat f_0(\xi)
    =
    \sum_{k=1}^{\infty}
    {a_k\over\Gamma(k)}\xi^{k-1}.
    \label{eq:spinBorelTransform}
\end{equation}
The nonlinear sectors in Eq.~\eqref{eq:fFullTransseries} give candidate Borel singularities at
\begin{equation}
    \xi_n=n\Xi,
    \qquad
    n=1,2,\ldots,
    \label{eq:spinBorelLocations}
\end{equation}
provided that the corresponding Stokes constants are nonzero. In particular, the leading singularity is located at
\begin{equation}
    \xi_1=\Xi={3\over2}\sqrt{1-16\lambda}.
    \label{eq:nearestSpinBorel}
\end{equation}
With the convention in Eq.~\eqref{eq:spinBorelTransform}, a sector of the form
\begin{equation}
    \e^{-n\Xi w}w^{n\beta+1}
\end{equation}
corresponds to a local Borel singularity
\begin{equation}
    \widehat f_0(\xi)
    \underset{\xi\to\xi_n}{\sim}
    {\mathcal{C}_n\over(\xi_n-\xi)^{\gamma_n}}
    \left[1+O(\xi_n-\xi)\right],
    \qquad
    \gamma_n=n\beta+2.
    \label{eq:spinBranchExponent}
\end{equation}
For generic values of $\lambda$, $\gamma_n$ is noninteger and the corresponding singularity is an algebraic branch point. At isolated values for which $\gamma_n$ becomes an integer, the local representation requires the corresponding resonant limiting form.

The singularity at $\xi=\Xi$ controls the leading large-order behavior
\begin{equation}
    a_k
    \sim
    \mathcal{K}_1
    {\Gamma(k+\beta+1)\over\Xi^{k+\beta+1}}
    \left[1+O(k^{-1})\right],
    \qquad
    k\to\infty,
    \label{eq:spinLargeOrder}
\end{equation}
where $\mathcal{K}_1$ contains the first Stokes constant together with the normalization of the first nonperturbative sector. Consequently,
\begin{equation}
    {a_{k+1}\over a_k}
    \sim
    {k+\beta+1\over\Xi}
    \left[1+O(k^{-1})\right].
    \label{eq:spinRatioTest}
\end{equation}

\subsection{Reconstruction of the repellor from the attractor asymptotic expansion}

The projective construction makes the resurgent relation between the dominant and subdominant linear modes transparent. 
For one fixed lateral canonical basis, define
\begin{equation}
    f^{(1)}(w)
    =
    \e^{-\Xi w}w^{\beta+1}\Phi_1^{(f)}(w).
    \label{eq:firstTwoSpinSectors}
\end{equation}
Across the positive real Borel axis, the leading Stokes connection of the formal dominant sector takes the form~\cite{Aniceto:2018bis,Dorigoni:2014hea}
\begin{equation}
    \mathfrak{S}_{0^+}f_{\rm dom}^{(0)}(w)
    =
    \mathfrak{S}_{0^-}f_{\rm dom}^{(0)}(w)
    +\mathfrak{s}_1\mathfrak{S}_{0^-}f^{(1)}(w)
    +O\!\left(\e^{-2\Xi w}w^{2\beta+1}\right),
    \label{eq:zeroOneStokesConnection}
\end{equation}
where $\mathfrak{s}_1$ is the Stokes constant associated with the leading Borel singularity.

At the level of the canonical projective family, the first nonperturbative sector satisfies
\begin{equation}
    \mathfrak{S}_{0^-}f^{(1)}(w)
    ={2w\over3}\mathcal{R}_{\rm dom}'(w)
    =\mathcal{R}_{\rm dom}(w)\left[f_-(w)-f_{\mathcal{S}_{\rm dom}}(w)\right],
    \label{eq:firstSectorBranchRelation}
\end{equation}
where $\mathcal{S}_{\rm dom}\equiv\mathcal{S}_{\rm dom}^{(-)}$ and $\mathcal{R}_{\rm dom}=\mathcal{S}_-/\mathcal{S}_{\rm dom}$.
Since $\mathcal{R}_{\rm dom}(w)\to0$ at late time,
\begin{equation}
    \mathcal{R}_{\rm dom}(w)
    =-
    \int_w^\infty {3\over2u}\mathfrak{S}_{0^-}f^{(1)}(u)\,du,
    \label{eq:RFromFirstSector}
\end{equation}
and hence
\begin{equation}
    f_-(w)
    =f_{\mathcal{S}_{\rm dom}}(w)
    -
    {\mathfrak{S}_{0^-}f^{(1)}(w)
    \over
    \displaystyle\int_w^\infty {3\over2u}\mathfrak{S}_{0^-}f^{(1)}(u)\,du}.
    \label{eq:repellerReconstruction}
\end{equation}
Thus the first nonperturbative sector associated with the leading Borel singularity contains the information needed to reconstruct the repellor. Because the exact attractor $f_+$ has the same formal dominant asymptotic series $f_{\rm dom}^{(0)}$, this resurgent information is equally encoded in the late time expansion of the regularity-selected attractor. The regularity condition fixes the nonperturbative completion of that expansion, but does not change its Borel singularity data.

Using Eq.~\eqref{eq:zeroOneStokesConnection}, the first sector may be extracted from the lateral discontinuity,
\begin{equation}
    \mathfrak{s}_1\mathfrak{S}_{0^-}f^{(1)}(w)
    =
    \mathfrak{S}_{0^+}f_{\rm dom}^{(0)}(w)
    -\mathfrak{S}_{0^-}f_{\rm dom}^{(0)}(w)
    +O\!\left(\e^{-2\Xi w}w^{2\beta+1}\right).
    \label{eq:firstSectorFromAttractorJump}
\end{equation}
The normalization-dependent Stokes constant cancels in the logarithmic reconstruction of the repellor.

Substituting Eq.~\eqref{eq:firstSectorFromAttractorJump} into
Eq.~\eqref{eq:repellerReconstruction}, the repellor can be expressed
directly in terms of the lateral discontinuity of the attractor
asymptotic expansion. To leading nonperturbative accuracy,
\begin{align}
    f_-(w)
    &=
    \mathfrak{S}_{0^-}f_{\rm dom}^{(0)}(w)
    -
    \frac{
        \mathfrak{S}_{0^+}f_{\rm dom}^{(0)}(w)
        -
        \mathfrak{S}_{0^-}f_{\rm dom}^{(0)}(w)
    }{
        \displaystyle
        \int_w^\infty {3\over2u}
        \left[
            \mathfrak{S}_{0^+}f_{\rm dom}^{(0)}(u)
            -
            \mathfrak{S}_{0^-}f_{\rm dom}^{(0)}(u)
        \right]du
    }
    +
    O\!\left(
        \e^{-\Xi w}w^{\beta+1}
    \right).
    \label{eq:repellerFromAttractorStokesData}
\end{align}
The normalization-dependent Stokes constant cancels between the
numerator and denominator. Thus, the resurgent data of the attractor
asymptotic expansion contain the information required to reconstruct
the repellor.

\subsection{Independent verification from the hypergeometric Borel transform}
\label{sec:hypergeometric_verification}

The Stokes connection can be verified directly from the exact Borel transform of the formal fluctuation series $\Phi_+(w)$. Substituting Eq.~\eqref{eq:Ssectors} into Eq.~\eqref{eq:linearSpinEquation} gives
\begin{equation}
    \Phi_\pm''
    +
    \left[
        \alpha_\pm+{2p_\pm+1\over w}
    \right]\Phi_\pm'
    +{p_\pm^2-9/4\over w^2}\Phi_\pm
    =0,
    \qquad
    \alpha_\pm=2r_\pm+{3\over2}=\pm\Xi.
    \label{eq:PhiEq}
\end{equation}
For
\begin{equation}
    \Phi_\pm(w)
    \sim
    \sum_{n=0}^{\infty}{c_n^{(\pm)}\over w^n},
    \qquad
    c_0^{(\pm)}=1,
\end{equation}
the coefficients satisfy
\begin{equation}
    c_n^{(\pm)}
    =
    {(n-1-p_\pm)^2-9/4\over\alpha_\pm n}
    c_{n-1}^{(\pm)}.
    \label{eq:recursion}
\end{equation}
Equivalently,
\begin{equation}
    c_n^{(\pm)}
    =
    {(a_\pm)_n(b_\pm)_n\over n!\,\alpha_\pm^n},
    \qquad
    a_\pm=-p_\pm-{3\over2},
    \qquad
    b_\pm=-p_\pm+{3\over2}.
    \label{eq:closedCoefficients}
\end{equation}
The Borel transforms are therefore
\begin{equation}
    \widehat{\Phi}_\pm(\zeta)
    =
    {}_2F_1\!\left(a_\pm,b_\pm;1;{\zeta\over\alpha_\pm}\right).
    \label{eq:BorelHypergeom}
\end{equation}
For the dominant formal sector, $\alpha_+=\Xi>0$ and the nearest positive-axis Borel singularity is at
\begin{equation}
    \zeta=\Xi=r_+-r_-.
    \label{eq:BorelSingularity}
\end{equation}
The parameters satisfy
\begin{equation}
    a_++b_+=1+\beta,
    \qquad
    1-a_+=b_-,
    \qquad
    1-b_+=a_-.
    \label{eq:hypergeomParameterRelations}
\end{equation}
For generic noninteger $\beta$, the singular part near $\zeta=\Xi$ is
\begin{align}
    \widehat{\Phi}_+^{\,\mathrm{sing}}(\zeta)
    &={\Gamma(\beta)\over\Gamma(a_+)\Gamma(b_+)}
    \left(1-{\zeta\over\Xi}\right)^{-\beta}
    {}_2F_1\!\left(
        a_-,b_-;1-\beta;
        1-{\zeta\over\Xi}
    \right).
    \label{eq:HypergeomSingularPart}
\end{align}
The discontinuity across the cut $\zeta>\Xi$ is consequently
\begin{align}
    \widehat{\Phi}_+(\zeta+i0)
    -
    \widehat{\Phi}_+(\zeta-i0)
    &=
    {2i\sin(\pi\beta)\Gamma(\beta)
    \over
    \Gamma(a_+)\Gamma(b_+)}
    \left(
        {\zeta-\Xi\over\Xi}
    \right)^{-\beta}
    {}_2F_1\!\left(
        a_-,b_-;1-\beta;
        -{\zeta-\Xi\over\Xi}
    \right).
    \label{eq:HypergeomLocalDiscontinuity}
\end{align}
At integer values of $\beta$, this expression is understood by
analytic continuation, with the corresponding resonant local
representation.

With the Borel convention appropriate to a series in $1/w$, the
regularized lateral resummations are defined by
\begin{equation}
    \mathfrak{S}_{0^\pm}\Phi_+(w)
    =
    w
    \int_0^{\e^{\pm i0}\infty}
    d\zeta\,
    \e^{-w\zeta}
    \widehat{\Phi}_+(\zeta).
    \label{eq:lateralSum}
\end{equation}
Since $\beta>2$ in the parameter range considered here, the branch
point is not locally integrable in the ordinary sense. The lateral
Laplace transforms in Eq.~\eqref{eq:lateralSum} are therefore
understood by analytic continuation, equivalently through a
regularized Hankel prescription around the branch point.

Applying the lateral Laplace transform to
Eq.~\eqref{eq:HypergeomLocalDiscontinuity} gives
\begin{equation}
    \mathfrak{S}_{0^+}\Phi_+(w)
    -
    \mathfrak{S}_{0^-}\Phi_+(w)
    =
    \mathcal{C}_{+-}^{(\Phi)}
    \e^{-\Xi w}w^\beta
    \mathfrak{S}_{0}\Phi_-(w),
    \label{eq:PhiJumpVerification}
\end{equation}
where
\begin{equation}
    \mathcal{C}_{+-}^{(\Phi)}
    =
    {2\pi i\,\Xi^\beta
    \over
    \Gamma(a_+)\Gamma(b_+)}.
    \label{eq:HypergeomStokesConstant}
\end{equation}
Here we use the convention
$\mathfrak{S}_{0^+}-\mathfrak{S}_{0^-}$; the opposite convention
reverses the overall sign.

The lateral sums of the full dominant formal sector define the two canonical functions $\mathcal{S}_{\rm dom}^{(\pm)}$,
\begin{equation}
    \mathcal{S}_{\rm dom}^{(\pm)}(w)
    =
    \e^{r_+w}w^{p_+}\mathfrak{S}_{0^\pm}\Phi_+(w),
    \label{eq:DfromLateralSum}
\end{equation}
with the normalization chosen above. Equation~\eqref{eq:PhiJumpVerification} therefore gives
\begin{equation}
    \mathcal{S}_{\rm dom}^{(+)}(w)-\mathcal{S}_{\rm dom}^{(-)}(w)
    =
    \mathcal{C}_{+-}^{(\Phi)}\mathcal{S}_-(w).
    \label{eq:FullJumpVerification}
\end{equation}
The repellor can be reconstructed directly from this Stokes jump,
\begin{equation}
    f_-(w)
    ={2w\over3}{d\over dw}
    \log\!\left[\mathcal{S}_{\rm dom}^{(+)}(w)-\mathcal{S}_{\rm dom}^{(-)}(w)\right],
    \label{eq:RepellerFromLinearJump}
\end{equation}
since the normalization-dependent Stokes constant drops out under the logarithmic derivative.

\subsection{Exact Whittaker completion and regularity selection}
\label{sec:whittaker_completion}

The exact Whittaker solution makes the relation between regularity and the Stokes parameter explicit. The standard Whittaker connection~\cite{NIST:DLMF} formula with $\mu=3/2$ gives
\begin{align}
    {1\over6}M_{\kappa,3/2}(z)
    &=
    {\e^{\pm i\pi(\kappa-2)}\over\Gamma(\kappa+2)}
    W_{\kappa,3/2}(z)
    +
    {\e^{\pm i\pi\kappa}\over\Gamma(2-\kappa)}
    W_{-\kappa,3/2}\!\left(\e^{\pm i\pi}z\right).
    \label{eq:WhittakerConnection}
\end{align}
Using the normalizations in Eqs.~\eqref{eq:ExactRegularSplus}, \eqref{eq:ExactSubdominantSminus}, and \eqref{eq:CanonicalDominantWhittaker}, this becomes
\begin{equation}
    \mathcal{S}_+(w)
    =\mathcal{S}_{\rm dom}^{(\pm)}(w)+\rho_{\rm att}^{(\pm)}\mathcal{S}_-(w),
    \label{eq:ExactRegularCompletion}
\end{equation}
where
\begin{equation}
    \rho_{\rm att}^{(\pm)}
    =
    \Xi^{2\kappa}
    {\Gamma(2-\kappa)\over\Gamma(2+\kappa)}
    \e^{\pm i\pi(\kappa-2)}.
    \label{eq:rhoAttExact}
\end{equation}
This is the explicit value of the fixed projective shift introduced in Eq.~\eqref{eq:rhoSigmaShift}. Therefore the early-time regularity condition fixes the nonperturbative transseries completion of the dominant late time series. In the regular basis the same statement is simply $\sigma=0$.

Since the exact regular solution is independent of the choice of lateral representation,
\begin{equation}
    \mathcal{S}_{\rm dom}^{(+)}(w)-\mathcal{S}_{\rm dom}^{(-)}(w)
    =
    \left[\rho_{\rm att}^{(-)}-\rho_{\rm att}^{(+)}\right]\mathcal{S}_-(w).
\end{equation}
Using the reflection formula for the Gamma function gives
\begin{equation}
    \rho_{\rm att}^{(-)}-\rho_{\rm att}^{(+)}
    =
    {2\pi i\,\Xi^{2\kappa}\over
    \Gamma(\kappa-1)\Gamma(\kappa+2)}.
    \label{eq:WhittakerStokesConstant}
\end{equation}
Finally,
\begin{equation}
    \beta=2\kappa,
    \qquad
    a_+=\kappa-1,
    \qquad
    b_+=\kappa+2,
\end{equation}
so that
\begin{equation}
    {2\pi i\,\Xi^{2\kappa}\over
    \Gamma(\kappa-1)\Gamma(\kappa+2)}
    =
    {2\pi i\,\Xi^\beta\over
    \Gamma(a_+)\Gamma(b_+)}
    =
    \mathcal{C}_{+-}^{(\Phi)}.
    \label{eq:StokesConstantsMatch}
\end{equation}
Thus the hypergeometric Borel discontinuity and the exact Whittaker connection formula describe the same Stokes phenomenon. At the same time, Eq.~\eqref{eq:rhoAttExact} shows that regularity fixes the otherwise invisible exponentially small completion. The exact attractor is therefore a unique projective solution in the regular basis, whereas the canonical late time transseries represents the same solution with a Stokes-dependent coordinate $\rho_{\rm att}^{(\pm)}$.

\section{Conclusion}
\label{sec:conclusion}

In this work, we have investigated the geometric and resurgent structure underlying a spin hydrodynamic attractor in Bjorken expansion. 
Our analysis starts from the second order linear equation governing the spin density and exploits the fact that the nonlinear attractor variable is defined through its logarithmic derivative. 
Using exact special function solutions, we show that early time regularity selects a unique attractor ray and fixes the corresponding nonperturbative completion of the late time transseries through a definite shift of the canonical transseries coordinate.

A projective structure is isolated from our analysis, identifying the nonlinear solution space with the projectivization of the two-dimensional linear solution space. 
A generic nonlinear solution depends only on the relative amplitude of the two exact solutions, which serves as a projective coordinate, with the attractor and repellor corresponding to two distinguished projective directions. 
The exact Whittaker solution further show that early time regularity uniquely selects the attractor branch, while the complementary exact solution is singular. 
At late time, generic solutions approach the same attractor because the repellor associated solution becomes subdominant. 
Thus the attractor selected by early time regularity coincides with the branch selected dynamically at late time.

We have further shown that the logarithmic derivative map converts the two linear modes into an infinite nonlinear transseries. 
Although this transseries contains sectors with actions given by integer multiples of the mode separation, these sectors do not represent independent degrees of freedom. 
They are generated algebraically by a single projective parameter and by the ratio of the two linear solutions. 
The same exponential separation determines the location of the leading Borel singularity of the perturbative expansion of the attractor and controls the factorial growth of its large order coefficients.
The normalized shear stress can likewise be formulated in a projective structure, although, unlike in the spin sector, the corresponding auxiliary linear solution space has no direct physical interpretation.

A central result of this work is that the resurgent data of the attractor encode the repellor branch. 
The first nonperturbative sector associated with the leading Borel singularity contains sufficient
information to reconstruct the ratio of the two linear modes and, consequently, the nonlinear repellor solution. 
This result was obtained directly from the Stokes connection between the perturbative and first nonperturbative sectors of the nonlinear transseries.
Besides, we also provided an independent analytic verification based the second order linear equation governing the spin density. 
The Borel transforms of the transseries associated with the two linear solutions can be expressed exactly in terms of hypergeometric functions. 
The connection formula across the leading Borel cut reproduces the exponential scale, algebraic prefactor, and complete fluctuation expansion of the subdominant linear mode. 
Under the logarithmic-derivative map, this Stokes connection yields a contribution associated with the repellor.
The two derivations therefore provide complementary descriptions of the same underlying resurgent structure.

Finally, the distinction between the regular projective basis and the canonical late time Stokes basis provides a unified view of these results.
The two are related by a fixed projective shift $\rho_\pm=\rho_{\rm att}^{(\pm)}+\sigma$.
The exact Whittaker connection formula determines $\rho_{\rm att}^{(\pm)}$ and reproduces the Stokes multiplier obtained from the hypergeometric Borel transform.
Thus regularity fixes the transseries completion of the attractor, while the Borel--Stokes discontinuity encodes the subdominant linear mode associated with the repellor.

\begin{acknowledgments}
This research is supported by the National Natural Science Foundation of China with Project Nos.~12635009.
\end{acknowledgments}

\bibliography{refs}
\end{document}